\documentclass[aps,prl,floats,twocolumn,superscriptaddress,showpacs]{revtex4}

\usepackage{graphicx,epsfig}
\usepackage{times}
\usepackage{graphics,dcolumn,bm,fleqn,epic,eepic,float}
\usepackage{amssymb,amsmath,multirow,rotate,color}

\begin{document}

\title{Entropy Rate of Diffusion Processes on Complex Networks}

\author{Jes\'us G\'omez-Garde\~nes}

\affiliation{Scuola Superiore di Catania, Via S. Paolo 73,
95123 Catania, Italy}

\affiliation{Institute for Biocomputation and Physics of Complex
Systems (BIFI), University of Zaragoza, Zaragoza 50009, Spain}

\author{ Vito Latora}

\affiliation{Dipartimento di Fisica e Astronomia, Universit\`a di Catania and
INFN, Via S. Sofia 64, 95123 Catania, Italy}

\date{\today}

\begin{abstract}
The concept of entropy rate for a dynamical process on a 
graph is introduced. 
We study diffusion processes where the node degrees are used as a local 
information by the random walkers.   
We describe analitically and numerically how the degree heterogeneity and correlations 
affect the diffusion entropy rate. 
In addition, the entropy rate is used to characterize complex 
networks from the real world. 
Our results point out how to design optimal diffusion processes that maximize 
the entropy for a given network structure, providing a new theoretical tool with 
applications to social, technological and communication networks.

\end{abstract}


\maketitle

Entropy is a key concept in statistical thermodynamics \cite{stat}, in the
theory of dynamical systems \cite{beck}, and in information theory
\cite{cover1991}.  In the realm of complex networks \cite{siam,bocca}, the
entropy has been used as a measure to characterize properties of the topology,
such as the {\em degree distribution} of a graph \cite{opt1}, or the {\em
shortest paths} between couples of nodes (with the main interest in
quantifying the information associated with locating specific addresses
\cite{rosvall}, or to send signals in the network \cite{trusina}.
Alternatively, various authors have studied the entropy associated with {\em
ensembles of graphs}, and provided, via the application of the maximum entropy
principle, the best prediction of network properties subject to the
constraints imposed by a given set of observations
\cite{bianconi,park,minnhagen}.  An approach of this type plays
the same role in networks as is played by the Boltzmann distribution in
statistical thermodynamics \cite{ stat}.

The main theoretical and empirical interest in the study of complex networks
is in understanding the relations between structure and
function. Besides, many of the interaction dynamics that takes place in
social, biological and technological systems can be analyzed in terms of {\em
diffusion processes} on top of complex networks, {\em e.g.}  data search and
routing, information and disease spreading \cite{siam,bocca}.
It is therefore of outmost importance to relate the properties of a diffusion
process with the structure of the underlying network.

In this Letter, we show how to associate an {\em entropy rate} to a
diffusion process on a graph.
In particular, we consider processes such as biased random walks on the graph
that can be represented as ergodic Markov chains. In this context,
the entropy rate is a quantity more similar to the Kolmogorov-\^{A}­Sinai
entropy rate of a dynamical system \cite{kolmo,latora}, than to the
entropy of a
statistical ensemble \cite{stat,siam}.  Differently from the network entropies
previously defined, the entropy rate of diffusion processes depends both on
the dynamical process (the kind of bias in the random walker) and
on the graph topology.
We provide the analytical expression that describes the entropy rate
in scale-free networks as a function of the bias in the walk,
and of the degree distribution and correlations.
We show how the values of the entropy rate can provide useful information
to characterize diffusion processes in real-world networks.
In particular, a maximum value of entropy is found for
different types of the bias in the diffusion processes, depending on the
network structure.

%
Let us consider a connected undirected graph with $N$ nodes (labelled
as $1, 2, ..., N $) and $K$ links, described by the adjacency matrix
$A=\{a_{ij}\}$
We limit our discussion to diffusion processes on the graph that
can be represented as {\em Markov chains} \cite{cover1991}. In
particular, we consider the case of biased random walks in which, at
each time step, the walker at node $i$ chooses one of the first
neighbors of $i$, let say $j$, with a probability proportional to the
power $\alpha$ ($\alpha \in \mathbb R$) of the degree $k_j$. Such
biased random walk corresponds to a time-invariant (the rule does not
change in time) Markov chain with a {\em transition probability
matrix} $\Pi$, with elements:
\begin{equation}
 \pi_{ji} = \frac{a_{ij} k_j^\alpha}{\sum_j a_{ij} k_j^\alpha }
\label{eq:biasrw_tm}
\end{equation}
Notice that $\Pi$ depends on either the graph topology and the kind of
stochastic process we are considering.  The exponent $\alpha$ allows to tune
the dependence of the diffusion process on the nodes' degree.  When
${\alpha}\neq 0$ we are introducing in the random movement of the particle a
bias towards high- ($\alpha >0$) or low-degree (when $\alpha <0$)
neighbors. On the other hand, when $\alpha=0$ the standard (unbiased) random
walk is recovered. Since the walker must move from a node to somewhere, we
have $\sum_{j} \pi_{ji} = 1$, thus $\Pi$ is a stochastic matrix.
If $w_i(t)$ is the probability that the random walker is at node $i$ at time
$t$ (with $\sum_{i=1}^N w_i(t)= 1 ~\forall t$), then the probability
$w_j(t+1)$ of its being at $j$ one step later is: $ w_j (t+1) = \sum_i
\pi_{ji} w_i(t)$.  Writing the probabilities $w_i(t)$ as a $N$-dimensional
column vector $ \mathbf{w}(t)= (~ w_1(t), w_2(t)\dots w_N(t)~)^\top$, the rule
of the walk can be expressed in matricial form as: $\mathbf{w} (t+1) =\Pi
\mathbf{w}(t)$.  In the case of an undirected and connected network, the
Perron-Frobenius theorem \cite{gantmacher59} assures that the dynamics
described by Eq.~(\ref{eq:biasrw_tm}) is an {\em ergodic Markov chain}
\cite{cover1991}. This means that the Markov chain has a unique {\em
stationary distribution} $\mathbf{w}^{*}$, such that $\lim_{t \rightarrow
\infty } \Pi^{t} \mathbf{w}(0) = \mathbf{w}^{*}$ for any initial distribution
$\mathbf{w}(0)$. In other words, any initial distribution of the random walker
over the nodes of the graph will converge, under the dynamics of
Eq.~(\ref{eq:biasrw_tm}), to the same distribution $\mathbf{w}^{*}$.

%
The dynamical properties of the above diffusion processes over the graph can
be accounted by evaluating the {\em entropy rate} of the associated Markov
chain that, in the case of an ergodic Markov chain, is given by
\cite{cover1991}:
\begin{equation}
   h= -\sum_{i,j}\pi_{ji}\cdot w^*_{i}\ln(\pi_{ji})
\label{eq:Entropy}
\end{equation}
The value of $h$ measures how the entropy of the biased random walk
grows with the number of hops.  This means that we can practically
represent the typical sequences of length $n$ generated by the
diffusion process by using approximately $n \cdot h$ information
units. In different words, $h$ measures the spreading
of a set of independent random walkers, in terms of number of
visited nodes.

To evaluate $h$ for a given graph we need to calculate the stationary
probability distribution $\mathbf{w}^{*}$. For this purpose,
we consider the probability
$W_{i\rightarrow j}(t)$ of going from node $i$ to node $j$ in $t$ time steps,
\begin{equation}
W_{i\rightarrow j}(t)=\sum_{j_1,j_2,...,j_{t-1}}\pi_{i,j_1}\cdot\pi_{j_1,j_2}
\cdot ...\cdot\pi_{j_{t-1},j}\;.
\end{equation}
Since the network is undirected we have $a_{ij}=a_{ji}$ $\forall i,j$.  Hence,
the relation between the two probabilities $W_{i\rightarrow j}(t)$ and
$W_{j\rightarrow i}(t)$ can be written as:
\begin{equation}
c_{i} k_{i}^{\alpha}W_{i\rightarrow j}(t)=c_{j} k_{j}^{\alpha}W_{j\rightarrow
i}(t)\;,
\end{equation}
where $c_{i}=\sum_{j}a_{ij}k_{j}^{\alpha}$. The above relation implies that
for the stationary distribution ${\bf w^{*}}$ the equation $c_i k_i^{\alpha}
w_j^*=c_j k_j^{\alpha} w_i^*$ holds, and hence ${\bf w^{*}}$ reads:
\begin{equation}
\label{eq:biasrw_stat}
w_i^*   = \frac{ c_{i} k_{i}^{\alpha}         }
               { \sum_l c_{l} k_{l}^{\alpha}  }
\;.
\end{equation}
By plugging expressions (\ref{eq:biasrw_tm})  and  (\ref{eq:biasrw_stat})
into the definition of entropy (\ref{eq:Entropy}), we finally get
a closed form for the entropy rate of degree-biased random walks on the graph:
\begin{equation}
 h=\frac{\sum_{i}k_{i}^{\alpha}\sum_{j}a_{ij}k_{j}^{\alpha}\ln(k_{j}^{\alpha})
-\sum_{i}k_{i}^{\alpha}c_{i}\ln(c_{i})}{\sum_{i}c_{i}k_{i}^{\alpha}}\;.
\label{eq:EntropyBRW}
\end{equation}
We notice that $h$ depends on the the kind of bias in the random walker
and also on the graph topology.
In the following, we first evaluate analytically the entropy rate
of unbiased and biased random walks on scale-free (SF) graphs with a
power-law degree distribution $P_k \sim k^{-\gamma}$, and
$\gamma>2$  \cite{siam,bocca}. 
Then, we study the entropy in networks from the real world.

%
{\
{\em{Unbiased Random Walks.-}}
In the particular case $\alpha =0$,
the transition probability reads $\pi_{ji}=a_{ij}/k_{i}$, and the stationary
distribution is easily obtained as: $ w_{i}^{*}=\frac{k_{i}}{2K} $.
Substituting this expression in Eq.~(\ref{eq:Entropy}) and changing the sum
over node indexes into a sum over degree classes, we can write the entropy
rate of a unbiased random walk on a network with degree distribution $P_k$ as:
\begin{equation}
h=\frac{N}{2K}\sum_{k} k P_k \ln(k)=\frac{\langle k\ln(k)\rangle}{\langle
  k\rangle}\;.
\label{eq:EntropyRW}
\end{equation}

In the case of SF networks of size $N$, the value of $h$ can be easily
expressed as a function of $\gamma$ and $N$ taking into account that the
maximum degree of the network is $k_{max}\sim k_0N^{1/(\gamma-1)}$, with $k_0$
being the minimum degree of a node. From Eq.~(\ref{eq:EntropyRW}), and
approximating $k$ as a continuum variable, we get:
\begin{equation}
h(\gamma,N)=\ln(k_0)+\frac{1}{\gamma-2}+
\frac{N^{\frac{2-\gamma}{\gamma-1}}\ln(N)}
{(\gamma-1)(N^{\frac{2-\gamma}{\gamma-1}}-1)}\;.
\label{eq:EntropyRWSF}
\end{equation}
The above expression diverges for SF networks when
$\gamma\rightarrow 2$. Conversely, when $\gamma>2$ SF
networks have a finite entropy in the thermodynamic limit:
$h(\gamma)=\ln(k_{0})+\frac{1}{\gamma-2}$.
\begin{figure}
\begin{center}
\epsfig{file=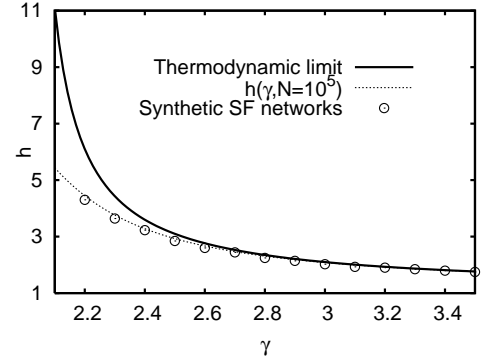,width=5.cm,angle=-90,clip=1}
\end{center}
\caption{{Entropy rate $h$ of unbiased random walks on SF networks
with $N=10^{5}$ nodes as a function of the exponent $\gamma$ of the degree
distribution. Numerical results (circles) are compared with the two analytical
curves corresponding to Eq.~(\ref{eq:EntropyRWSF}) (dashed line)
and to the limit $N \rightarrow \infty$ (solid line). }}
\label{fig:1}
\end{figure}

In order to check the analytical results we have constructed ensembles of
$10^2$ SF networks with $N=10^5$ nodes and different values of
$\gamma$. We have obtained numerically the stationary distribution ${\bf
w^{*}}$, and computed the entropy directly from Eq.~(\ref{eq:Entropy}).  The
results, averaged over the ensemble of networks, are reported in
Fig.~\ref{fig:1} as a function of $\gamma$. We notice a good agreement between
numerics and Eq.~(\ref{eq:EntropyRWSF}).

%
{\em{Biased Random Walks.-}}
Let us now concentrate on degree-biased diffusion ($\alpha \neq 0$). In this
case, the entropy rate of Eq.~(\ref{eq:EntropyBRW}) can be re-written by
changing again the sums over node indexes into sums over degree classes, as:

\begin{equation}
h = -\frac{\sum_{k}k^{\alpha} P_{k}\left(C_{k}\ln({C}_{k})
- \sum_{k^{'}}\alpha
{k^{'}}^{\alpha} P_{k^{'},k}
\ln(k^{'})\right)}{\sum_{k} {C}_{k}k^{\alpha} P_{k}  }
\label{eq:EntropyDBRW1}
\end{equation}
\medskip
where $C_{k}=k\sum_{k^{'}} {k^{'}}^{\alpha} P_{k^{'},k} $, and $P_{k^{'},k}$ is
the conditional probability that a link from a node of degree $k$ ends in a
node with degree $k^{'}$. We notice that the entropy rate of biased random
walks depends on the degree distribution of the network, $P_k$, and on the
conditional probabilities $P_{k^{'},k}$. In the particular case of a
network with no degree-degree correlations we
can write $P_{k^{'},k}=kP_{k}/\langle k\rangle$, and the expression for the
entropy reduces to:
\begin{equation}
h=(1-\alpha)\frac{\langle k^{\alpha+1}\ln(k)\rangle}{\langle
  k^{\alpha+1}\rangle}+\ln\left(\frac{\langle k^{\alpha+1}\rangle}{\langle
  k\rangle}\right)\;.
\label{eq:EntropyDBRW2}
\end{equation}
This expression only depends on the degree distribution of the network.
For SF networks, we get in the the continuum-degree
approximation:
\begin{eqnarray}
h(\gamma,\alpha,N)&=&\frac{1-\alpha}{\gamma-\alpha-2}+
\frac{(1-\alpha)N^{\frac{\alpha+2-\gamma}{\gamma-1}}\ln(N)}
{(\gamma-1)(N^{\frac{\alpha+2-\gamma}{\gamma-1}}-1)}
\nonumber
\\
&+&\ln\left[\frac{k_0(\gamma-2)(N^{\frac{\alpha+2-\gamma}{\gamma-1}}-1)}
{(\gamma-\alpha-2)(N^{\frac{2-\gamma}{\gamma-1}}-1)}\right]\;.
\label{eq:EntropySF-DBRW-N}
\end{eqnarray}
When $N \rightarrow \infty$, the entropy rate in SF networks
with $\gamma<2+\alpha$ diverges as $h\sim \ln(N)$.
On the other hand, when $\gamma>2+\alpha$, the entropy rate in the
limit $N \rightarrow \infty$ is finite and equal to:
\begin{equation}
h(\gamma,\alpha)=\frac{1-\alpha}{\gamma-\alpha-2}+
\ln\left[\frac{k_0(\gamma-2)}{\gamma-\alpha-2}\right]\;.
\label{eq:EntropySF-DBRW}
\end{equation}
Such an expression, valid in infinite size limit, shows a monotone growth of
the entropy $h(\gamma,\alpha)$ with the degree-bias $\alpha$, with $h$ tending
to infinity as $\alpha\rightarrow(\gamma-2)^-$.
More interestingly, the entropy rate in finite networks,
Eq. (\ref{eq:EntropySF-DBRW-N}), shows a single maximum at a value of $\alpha$
that depends on $\gamma$. This result is a consequence of the
interplay between diffusion process and network topology. It indicates
that, for a given network, is possible to maximize the entropy
of the process by opportunely tuning the bias $\alpha$ of the
walker.

\begin{figure}
\begin{center}
\epsfig{file=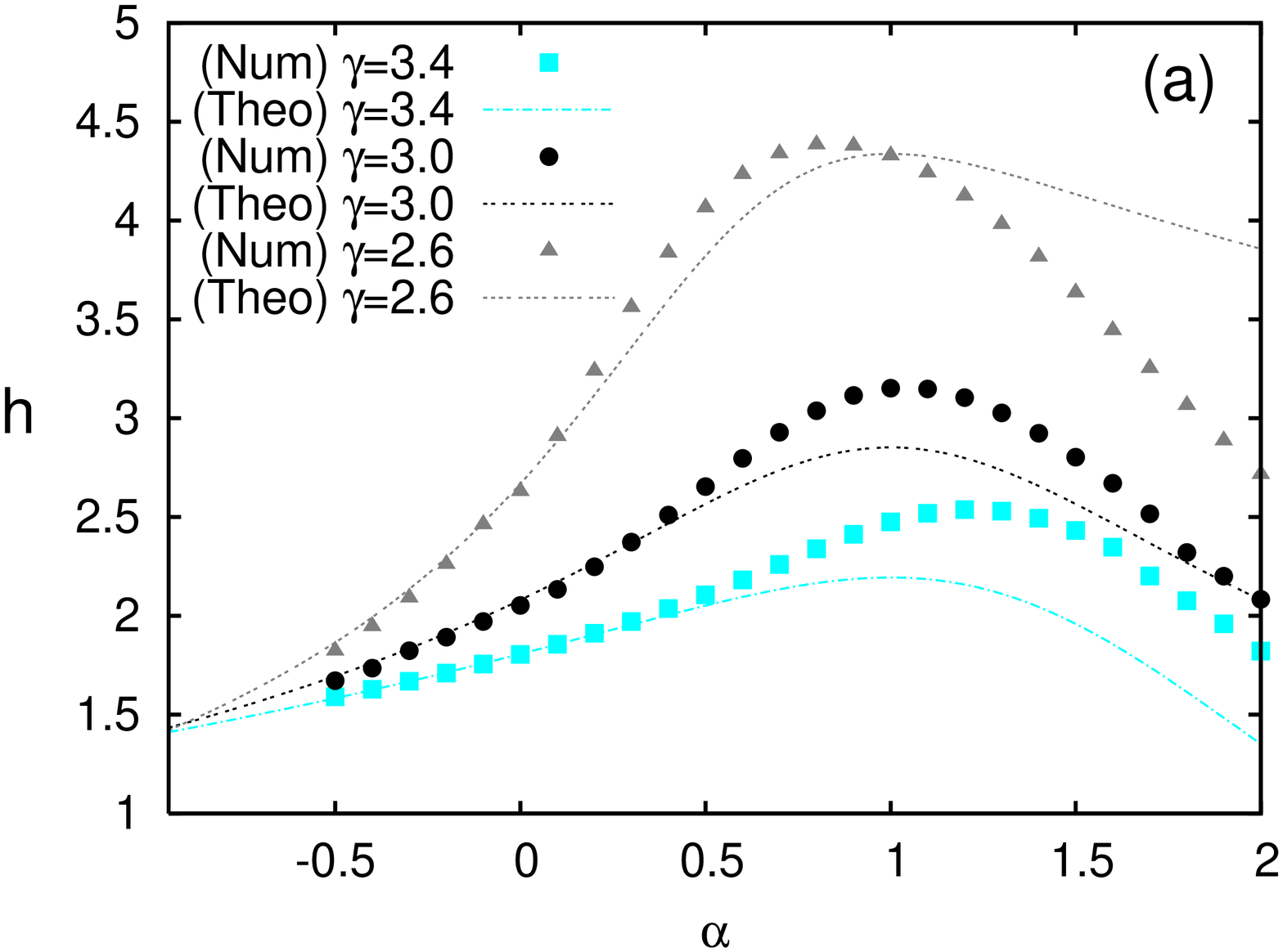,width=7.7cm,angle=-0,clip=1}
\\
\epsfig{file=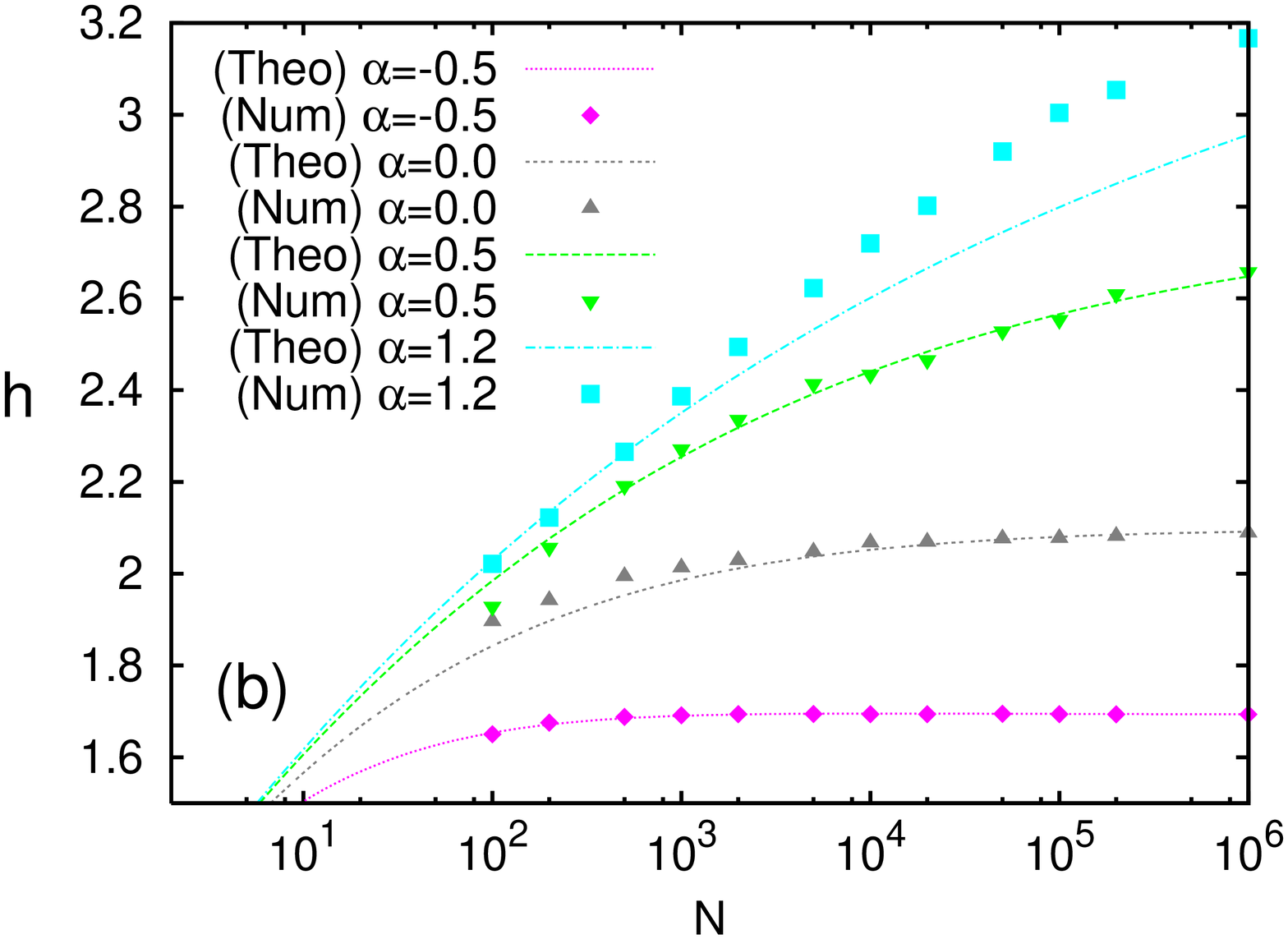,width=7.7cm,angle=-0,clip=1}
\end{center}
\caption{(color online). {(a)} Entropy rate $h$, as a function of $\alpha$,
for $\alpha$-biased random walks on SF networks with $N=10^5$ nodes and
$\gamma=2.6, 3, 3.4$. Symbols represent the values of $h$ found numerically,
while the lines are the corresponding analytical predictions
$h(\gamma,\alpha,N)$ of Eq.~(\ref{eq:EntropySF-DBRW-N}).  {\bf (b)} Entropy
rate $h$ for $\alpha$-biased random walks on SF networks with $\gamma=3$, as a
function of the system size $N$ and for several values of $\alpha$.  Again,
symbols are the results of numerical simulations, while the lines correspond
to Eq.~(\ref{eq:EntropySF-DBRW-N}).}
\label{fig:2}
\end{figure}

To check the above analytical expressions we have computed numerically the
entropy rate of degree-biased random walkers on computer-generated
uncorrelated SF networks, as we did for the unbiased case.
In Fig.~\ref{fig:2}.a we report the entropy rate as a function of the degree
bias $\alpha$ for SF networks of size $N=10^{5}$. In
Fig.~\ref{fig:2}.b we show the scaling of $h$ with the system size $N$, in
SF network with $\gamma=3$.  In both cases
Eq.~(\ref{eq:EntropySF-DBRW-N}) is in good agreement with the numerical
results reproducing the qualitative behavior of $h$ as a function of $\alpha$
(being the global maximum of $h$ well reproduced) and $N$ (being both the
divergence of $h$, for $\alpha>\gamma-2$, and the asymptotic finite value
of $h$, when $\alpha<\gamma-2$, correctly reproduced) \cite{NotaErr}.

%
{\em{Real Networks.-}}
Up to now, we focused on the entropy rate of biased random walks on SF
networks. However, real networks are not perfect scale-free and, more
importantly, show additional important structural properties such as
degree-degree correlations, motifs and
community structures \cite{siam,bocca}.
Now we propose to characterize a real network by studying different
diffusion processes on top of it, and finding the optimal value of
the bias that maximizes the entropy.
As reference system, we compare the entropy rate $h$ of $\alpha$-biased
random walks on the network, with the entropy rate $h^{Rand}$ obtained, from
Eq.~(\ref{eq:EntropyDBRW2}), for a randomized version of the network, with the
same degree sequence of the real one \cite{sneppen}.
For this purpose, we have analyzed $10$ different networks reported in Table
\ref{tab:1}, corresponding to thre different functional classes where
diffusion of data, rumors, viruses and diseases, takes place, namely
{\em (i)} transportation, {\em (ii)} technological/communication and {\em
(iii)} social networks.

\begin{table}
  \caption{Properties of the $10$ real networks analyzed. $N$ is the number of
  nodes in the giant connected component, $\langle k\rangle$ is the average
  degree. The ratio of entropy rates, $h/h^{Rand}$, is reported for a linear
  ($\alpha=1$) degree-biased random walk. Finally we report the optimal value
  of  $\alpha$ that maximizes $h$ for each network.\label{tab:1}}
\begin{center}
\begin{tabular}{cc|c|c|c|c}
 \hline
 \hline
\textbf{ network } & \textbf{ ref. } & \textbf{$N$} & \textbf{$\langle
  k\rangle$} & \textbf{$ h/h^{Rand}$} &  \textbf{$\alpha^{opt}$}\\
\hline
U.S. Airports & \cite{AIR} & 500 & 11.92 & 0.964 & 0.8 \\
  \hline
Internet routers & \cite{AS} & 228263 & 2.80 & 1.191 & 1.7 \\
Internet A.S. & \cite{AS} & 1174 & 4.19 & 0.662 & 0.6\\
WWW & \cite{WWW} & 325729 & 6.70 & 0.867 & 0.9 \\
P2P & \cite{YamirFang} & 79939 & 4.13 & 0.613 & 0.7\\
\hline
%
Sci. Coll. (cond-mat) & \cite{SCN} & 12722 & 6.28 & 1.091 & 1.5 \\
Sci. Coll. (astro-ph) & \cite{Cardillo} & 13259 & 18.62 & 1.071 & 1.5\\
U.S. patents & \cite{Patents} & 230686 & 4.81 & 1.113 & 1.5 \\
E-mail & \cite{Mail} & 1133 & 9.62 & 1.019 & 1.2\\
P.G.P & \cite{PGP2} & 10680 & 4.56 & 1.176 & 1.3\\
  \hline
  \hline
\end{tabular}
\end{center}
\end{table}

In Fig.~\ref{fig:3} we report, for six of the networks, the results obtained
as a function of the bias parameter ${\alpha}$.
Two different behaviors emerge clearly for $\alpha>0$ \cite{Nota},
namely the entropy of the real network $h$ is either larger or smaller than
$h^{rand}$ for all the range of positive values of $\alpha$. In table
\ref{tab:1} we summarize this result by reporting the ratio $h/h^{Rand}$ for
$\alpha=1$ (linear bias). We found that social networks
have always $h>h^{Rand}$, while the
other networks have $h < h^{Rand}$, with the exception of Internet
routers. This difference in the entropy rate has
its roots mainly on the different types of degree-degree correlations of the
network, and points out that assortativity facilitates
the spread of the diffusion.
The optimal degree-bias, $\alpha^{opt}$ that produces the
maximal entropy rate is also reported in Table \ref{tab:1}. The
results indicate that for assortative
networks ({\em e.g.} social networks) the  maximal entropy rate is
obtained with a super-linear diffusion, while for dissasortative
networks $\alpha^{opt}$ is located in the sub-linear bias region.

\begin{figure}
\begin{center}
\epsfig{file=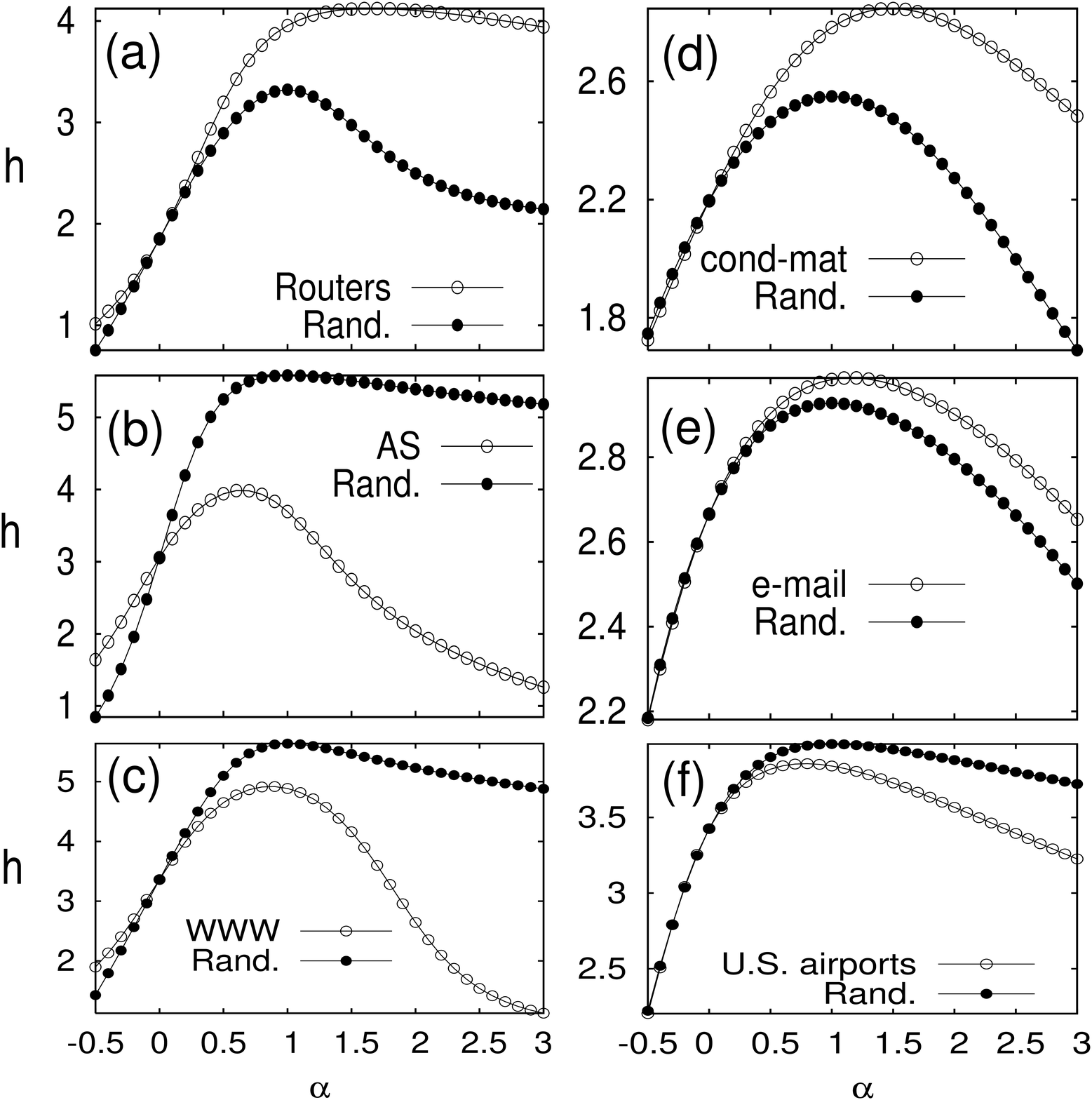,width=7.5cm,angle=-0,clip=1}
\end{center}
\caption{{Entropy rate $h$ for $\alpha$-biased random walks on six of the
networks in Table \ref{tab:1} (filled circles).  Such entropy rate is compared
with that obtained for the randomized version of the network (full
circles). Both entropy rates are shown as a function of the degree-bias
parameter ${\alpha}$.}}
\label{fig:3}
\end{figure}

%
Summing up, in this Letter, we have introduced the entropy rate of
degree-biased random walks on networks, a measure that is
particularly suited to capture the interplay between network
structure and diffusion dynamics.
We have studied the dependence of
the entropy rate with the topology of synthetic and real
networks, in particular with the
heterogeneity of degree distributions and the nature of the
degree-degree correlations. The results indicate how it is possible
to tune the bias in the random walk in order to maximize the
entropy rate on a given topology.
The method introduced can find useful applications to cases where diffusion
in complex networks is the mechanism at work, such as
in the search of efficient algorithms for data search in the WWW,  in
the improvement of information dissemination in social
networks, or in the design of large impact virus/antivirus spreading
in computer networks.
The approach adopted here can be easily generalized to other types
of diffusion processes and to more general network topologies, such as
weighted graphs and also, with some appropriate modifications to directed
and unconnected graphs.

\begin{acknowledgments}
The authors are indebted to M. Chavez and H. Touchette for many helpful
discussions on the subject.  
\end{acknowledgments}


\begin{thebibliography}{99}
\bibitem{NotaErr} Note that the effect of the continuum-degree approximation,
made in the calculation, become more evident as $\alpha$ grows, because for
large $\alpha$ the discreteness effects in $k$ are amplified.
\bibitem{Nota} Note that for $\alpha=0$ both the real network and the
  randomized ensemble have the same value of $h$ since in this limit Eq.~(\ref{eq:EntropyDBRW2}) (or alternatively Eq.~(\ref{eq:EntropyRW})) is exact
  and it only depends on the degree sequence of the network.
\end{thebibliography}
\end{document}